\documentclass[nolinenumbers]{aastex631}

\shorttitle{Rapid Tidal Evolution of Enceladus}
\shortauthors{\'Cuk and El Moutamid}
\graphicspath{{./}{figures/}}

\begin{document}

\title{A Past Episode of Rapid Tidal Evolution of Enceladus?}

\correspondingauthor{Matija \'Cuk}
\email{mcuk@seti.org}

\author[0000-0003-1226-7960]{Matija \'Cuk}
\affiliation{SETI Institute \\
339 N Bernardo Ave\\
Mountain View, CA 94043, USA}

\author[0000-0002-4416-8011]{Maryame El Moutamid}
\affiliation{Cornell Center of Astrophysics and Planetary Sciences\\ Department of Astronomy and Carl Sagan Institute\\
Cornell University\\
326 Space Science Building\\
Ithaca, NY 14853, USA}

\begin{abstract}

Saturn possesses a dynamically rich system containing numerous moons and impressive rings. Whether the rings of Saturn are much younger than the planet itself has been a long-open question; more recently a young age has been proposed for some moons. Recent detection of the fast orbital evolution of Rhea and Titan strongly suggest a highly frequency-dependent tidal response of Saturn, possibly through excitation of inertial waves within the planet's convective envelope. Here we show that the resonance locking to inertial waves cannot explain the dynamical structure of the Saturnian system or the current tidal heating of Enceladus. On the other hand, both the observation and our modelling results indicate that the system is not consistent with evolution under equilibrium tides. We propose that the system's architecture can best be explained by relatively high ``background'' tidal response coupled with discrete resonant modes. In this view, only Titan may be in a true long-term resonance lock with a tidal mode of Saturn. Rhea is most likely currently experiencing a transient period of fast tidal evolution as it passes through a mode, rather than being locked to it. Assuming that Enceladus went through a temporary period of fast tidal evolution, we can reproduce its present resonance with Dione and satisfy other dynamical constraints. Additionally, we conclude that the long-term tidal response of Saturn to Tethys must be weaker than expected from frequency-independent tides, as already found by observations. 

\end{abstract}

\keywords{Saturnian Satellites (1427) --- Celestial mechanics (211) --- Orbital resonances(1181) --- N-body simulations (1083)}

\section{Introduction} \label{sec:intro}

Before the Cassini mission to Saturn (2004-2017) it was widely thought that the major moons of Saturn (Mimas and larger) are as old as the planet, and that the moons' orbital evolution is driven by equilibrium tides within Saturn. ``Equilibrium tides'' refers to classical tidal theory described by textbooks like \citet{md99}, in which there is little or no dependence of tidal dissipation on the orbital frequency of the perturbing moon. These two assumptions constrained the tidal quality factor $Q$ which quantifies tidal dissipation within Saturn to $Q > 18,000$ (NB: lower $Q$ means higher dissipation), or else Mimas would have been within the rings less than 4.5 Gyr ago \citep{md99}.


The detection of thermal flux of 10-15 GW on Enceladus \citep{por06, how11} challenged this estimate of the tidal evolution rate. \citet{mey07} showed that, {\it assuming an equilibrium state}, tidal heating of Enceladus through its orbital resonance with Dione produces $(18,000 / Q) \times 1.1$~GW. Their result implies that either the tidal $Q$ of Saturn is an order of magnitude lower than $18,000$, or the heating of Enceladus is not in equilibrium. The latter explanation of the observations was initially dominant, but the astrometric work of \citet{lai12} suggested a migration rate of most major moons that is an order of magnitude faster than previously estimated. A notable implication of the results of  \citet{lai12} was that the tidal $Q \approx 1700$, and the tidal heating of Enceladus could be in equilibrium.   

\citet{cuk16}, assuming equilibrium-type tides and a constant tidal $Q \simeq 1700$ as found by \citet{lai12}, analyzed the orbital histories of the three largest moons interior to Titan: Tethys, Dione and Rhea. \citet{cuk16} found that their orbits are consistent with Dione and Rhea crossing their mutual 5:3 mean-motion resonance (MMR) in the past. \citet{cuk16} also modeled past passage of Tethys and Dione through their 3:2 MMR and found that this event excites inclinations well above the observed values, implying  that this resonance passage did not happen. Assuming $Q=1700$, this relative dynamical age (i.e. Dione-Rhea 5:3 MMR was crossed in the past, but Tethys-Dione 3:2 MMR was not) translates to an absolute age of the system below about 100 Myr. While the age of Saturn's rings is hotly debated, this result is consistent with the estimate of the rings' age derived by \citet{cuz98}. \citet{cuk16} concluded that the rings and moons interior to Titan formed in a dynamical instability about 100 Myr ago, in which the previous generation of moons was disrupted in collisions, and then largely re-accreted into the observed satellites.    

Some of the more recent findings are consistent with a young rings and satellite system. The relatively small mass \citep{ies19} and rapid evolution \citep{odo19} of the rings suggest a relatively young age, but other authors argue that ring pollutants are lost faster than ice, implying an older age \citep{cri19}. It appears that dominant past impactors in the Saturnian system are different from Kuiper Belt objects \citep{zah03, sin19}, and are possibly planetocentric \citep{fer20, fer22a, fer22b}. This would be consistent with, but would not require, a recent origin of the system. More recently, \citet{wis22} proposed an alternate proposal for a recent cataclysm that originates not in the inner system but in an instability between Titan and a past resonant moon \citep[cf. ][]{asp13, ham13}; the full consequence of this scenario for the inner moons is still unclear.

Since the work done by \citet{cuk16}, analysis of the moons' motions using both Earth-based astrometry and Cassini data \citep{lai17, lai20, jac22} has strongly suggested that the tidal evolution in the Saturnian system is not driven by equilibrium tides. Rhea has been found to migrate outward many times faster than predicted by equilibrium tides, with an orbital evolution timescale of $a/\dot{a} = 6~Gyr$. Additionally, \citet{lai12} find that Titan is migrating with a 11-Gyr timescale, but that is disputed by \citet{jac22}, who find a $>$100~Gyr timescale for Titan's evolution. Both groups agree on the fast evolution of Rhea, ruling out equilibrium tides as the only form of tidal dissipation within Saturn. 

The pattern of tidal evolution found by \citet{lai20} matches the expectations of resonant locking theory of \citet{ful16}. In this theory, Saturn's tidal response is exceptionally high at certain synodic frequencies, at which the tidal perturbations from the moons are resonant with internal oscillation modes of the planet. If the planet's structure were static, the moons would just quickly evolve through these frequencies without much consequence for their long-term evolution. However, due to changes in the resonant frequencies of the planet, the orbital locations at which a moon is resonant with the planet move outward, pushing the moons along faster than they would move through equilibrium tides \citep{ful16}. If this is true, the moon is evolving due to a ``resonance lock'' with the planet's interior. Similar migration timescales for multiple moons imply that inertial waves within the planet are the type of oscillation driving the evolution. \citet{lai20} suggest that the rate of tidal evolution for all moons at any time in their history may be given as $\dot{a}/a=(3 t_s)^{-1}$, where $t_s$ is the age of the planet at any time.

Orbital evolution of all of Saturn's moons with a uniform relative migration rate rules out mutual MMR crossings; while approximately equal, but not identical, evolution rates would greatly extend the time between any MMR crossings. If this is the mechanism behind the evolution of the Saturnian system, the constraints on its age proposed by \citet{cuk16} on the basis of MMRs do not apply. However, MMRs currently present in the system, as well as secular resonances, still need to be reconciled with this hypothesis of parallel orbital evolution.



Equilibrium tides that have previously been assumed to dominate orbital evolution of Saturnian satellites have a strong dependence on orbital distance, $\dot{a}/a \propto a^{-6.5}$ \citep{md99}. On the other hand, resonant lock produces orbital evolution that is either independent of orbital distance, $\dot{a}/a \ne f(a)$, in case of locking to inertial waves, or is faster for more distant satellites, $\dot{a}/a \propto a^{3/2}$, in case of locking to Saturn-frame modes \citep{ful16, lai20}.  Therefore, it is likely that equilibrium tides would dominate evolution in a zone closer to the planet, while the resonance lock (if present) would dominate evolution of more distant satellites. The exact boundary between these two regimes depends on the strength of the planet's tidal response and the rate at which resonant peaks in tidal response are shifting. Note that the equilibrium tidal evolution rate is directly proportional to the satellite's mass, while resonance lock evolution rate is independent from it, so the distance at which each of the mechanisms dominates will not be the same for moons of all sizes. Additionally, in the innermost zone close to the Saturn's rings, ring-torque driven evolution \citep{gol79} may dominate over all types of tidal evolution. 

Here we will consider several specific dynamical features of the Saturnian system in the light of potential mechanisms of tidal evolution

\section{Constraints on Saturn's Tidal Response from Current Tidal Heating of Enceladus}\label{sec:heat}

Enceladus and Dione are currently locked in a MMR with an argument $2\lambda_D-\lambda_E-\varpi_E$, where $\lambda$ and $\varpi$ are the mean longitude and the longitude of pericenter, respectively \citep{md99}, while subscripts D and E refer to Dione and Enceladus. This resonance keeps the moons' orbital period ratio close to 1:2, and acting to increase the eccentricity of Enceladus over time. Satellite tides within Enceladus act in the opposite direction, damping Enceladus's eccentricity and releasing heat in the process. This tidal dissipation within Enceladus is thought to drive the observed geological activity on Enceladus \citep{por06}. 

The amount of heat observed on Enceladus \citep{how11}, combined with a relatively low eccentricity ($e_E=0.005$), implies that Enceladus is very dissipative with tidal parameters given by $Q/k_2 \approx 100$ (or even smaller if there is more undetected heat), consistent with a fluid response to tides \citep{lai12}. Given that this eccentricity would damp in under 1~Myr, it is natural to assume that the eccentricity of Enceladus is continuously re-excited by the resonance with Dione, so that the eccentricity and the tidal heating are in a long-term equilibrium. In context of equilibrium tides, this would require $Q \approx 1700$ for Saturn \citep{mey08, lai12}. The dynamics of the tidal heating of Enceladus in the resonance lock scenario is yet to be fully modeled.

For the purposes of a MMR between two moons, there are several possible configurations of resonant modes. The simplest case is that only the inner moon is locked with a resonant mode, with the outer moon evolving only due to interaction with the inner. Alternatively, two moons could be locked to two converging modes. Convergent modes, however, are not compatible with a moon-moon resonance, as the inner moon would push the outer moon outward away from its mode, and we are back to the original case of only the inner moon evolving through resonant locking. A third possibility are simultaneous divergent resonant modes, but they would preclude capture into the moon-moon mean-motion resonance, which requires convergent evolution \citep{md99}. The fourth possibility is also conceivable: two modes evolving in parallel, with the inner moon experiencing stronger torque by being closer to the mode center, while the outer moon is subject to outward acceleration, both from the resonant mode and the perturbations of the inner moon. While this configuration is in principle consistent with a MMR, it is clear that it is sensitively dependent on the moons' locations relative to the resonant modes, and that the stability of this state would need to be investigated in greater detail. Therefore, here we will concentrate on the scenario in which only the inner moon is in resonance lock.

It can be shown that a MMR driven by resonance locking of the inner moon alone to inertial modes is not consistent with the present heating rate of Enceladus (assuming an equilibrium). \citet{mey07} show that the power of tidal heating of the inner moon of a resonant pair, assuming no tidal torque on the outer moon and all eccentricities being in equilibrium, is:
\begin{equation}
H = n_E T_E - {T_E \over L_E+L_D} \left( {G m_S m_E \over a_E} + {G m_S m_D \over a_D} \right) 
\label{meyer}
\end{equation}  
where $T_E$ is the tidal torque on Enceladus, while $L$ is angular momentum, $n$ is mean motion, $a$ is semimajor axis, $m$ is mass, and $G$ is gravitational constant.  Given that the moons are evolving in parallel due to mean-motion resonance lock, the torque has a simple relation to the evolution rate \citep{mey07}:
\begin{equation}
T_E={1 \over 2}{\dot{a} \over a} \left( {L_E + L_D}\right) = {1 \over 2 t_a} \left( {L_E + L_D}\right)
\label{torque}
\end{equation}
where $t_a$ is the evolution timescale. Substituting this back into Eq. \ref{meyer}, and assuming that the angular momentum and energy of Dione are much bigger than those of Enceladus (accurate to about 10\%), we get
\begin{equation}
H \approx {1 \over 2 t_a} \left( n_E L_D - {G m_S m_D \over a_D} \right) 
\label{meyer2}
\end{equation} 
Recognizing that $G m_S/a_D^3=n_D^2$, $n_E \approx 2 n_D$ and, for small eccentricities, $L_D = m_D a_D^2 n_D$, we get:
\begin{equation}
H \approx {G m_S m_D \over 2 t_a a_D} = {5 \times 10^{28} \ {\rm J} \over t_a} 
\label{heat}
\end{equation}
If we assume $t_a=9$~Gyr, as proposed by \citet{lai20} as being typical value for orbital evolution due to resonance locking in the Saturnian system, $H \approx 180$~GW, more than an order of magnitude in excess of the observed value \citep{how11}. Therefore it appears that the observed resonance between Enceladus and Dione cannot be maintained by only Enceladus being locked to an inertial wave within Saturn.

On the other hand, if the resonance lock evolves uniformly in a reference frame rotating with Saturn, with the dominant semi-diurnal tidal period changing by a constant absolute rate for all satellites \citep{ful16}, steady-state heating of Enceladus will be much lower. The Saturn-frame resonance lock would produce $a/\dot{a} \approx 200$~Gyr for Enceladus, based on Titan's evolution timescale of 11~Gyr \citep{lai20} and the $\dot{a}/a \propto a^{3/2}$ \citep{ful16}. This rate of evolution gives us a steady-state heat flow of about 8~GW for Enceladus, close but below the measured value (which is by itself the lower limit of the real flux). One caveat with this calculation is that we assumed no contribution from equilibrium tides acting on Enceladus and (more importantly) Dione, meaning that Saturn's ``background'' $Q/k_2 \gg 30,000$, or otherwise the tidal heating would be even lower. 

If we follow recent results of \citet{jac22} who find a possible resonant lock for Rhea but not Titan, the timescale for Enceladus's evolution through resonant lock (assuming divergent modes anchored at Rhea) could be in the 30-40~Gyr range, with the associated tidal heating in the 40-55~GW range. Models of Enceladus's ice shell suggest a heat loss rate in the range 25-40~GW  \citep{hem19}, and the total tidal heating rate of Enceladus likely exceeds the measured value (possibly due to distributed tidal heating outside the South Polar Terrain), so there may not be a discrepancy between these predictions and the actual heat production rate.

If we assume equilibrium tides and a steady-state, then a $Q/k_2=5000$ can explain the observed heat flux of Enceladus, as suggested by \citet{lai12}. Note that a combination of inertial wave lock and equilibrium tides can also explain the observed flux, as long as the equilibrium tidal evolution rate of Dione is about 90\% of the drift rate of the inertial mode Enceladus is locked to. This complex (but possible) setup still gives us about the same ``background'' $Q/k_2=5000$ as the assumption of pure equilibrium tides. The possibility of Enceladus and Dione being locked to two normal modes evolving almost in unison is also conceivable (and would result in lower equilibrium heat flows), but we assess it as less likely.

\section{Constraints on Saturn's Tidal Response from Orbital Resonances}

\subsection{Past 2:1 MMR between the Horseshoe Moons and Enceladus?}\label{ssec:janus}

Janus and Epimetheus are inner moons of Saturn that are currently in a horseshoe orbital configuration which undergoes a reversal every four years \citep{md99}. Gravitational interactions with Saturn's rings make the width of the ``horseshoe'' decrease over time, and the timescale for this evolution is on the order of $10^7$~yr \citep{lis85}. For our purposes Janus and Epimetheus (and their presumed parent body) are interesting because they are expected to migrate very fast due to ring torques, potentially crossing resonances with other moons. \citet{taj17} estimate that Janus migrates about 40 km/Myr due to ring torques, and the two moons' putative parent body (if they originated in a breakup) would have migrated even faster, at 50 km/Myr. Janus and Epimetheus are about 1500~km exterior to the 2:1 mean-motion resonance with Enceladus. Ignoring the tidal evolution of Enceladus for the moment, it appears that the precursor of Janus and Epimetheus should have crossed this resonance about 30 Myr ago, putting constraints on the age and evolution of the system.

\begin{figure*}
\epsscale{1.2}
\plotone{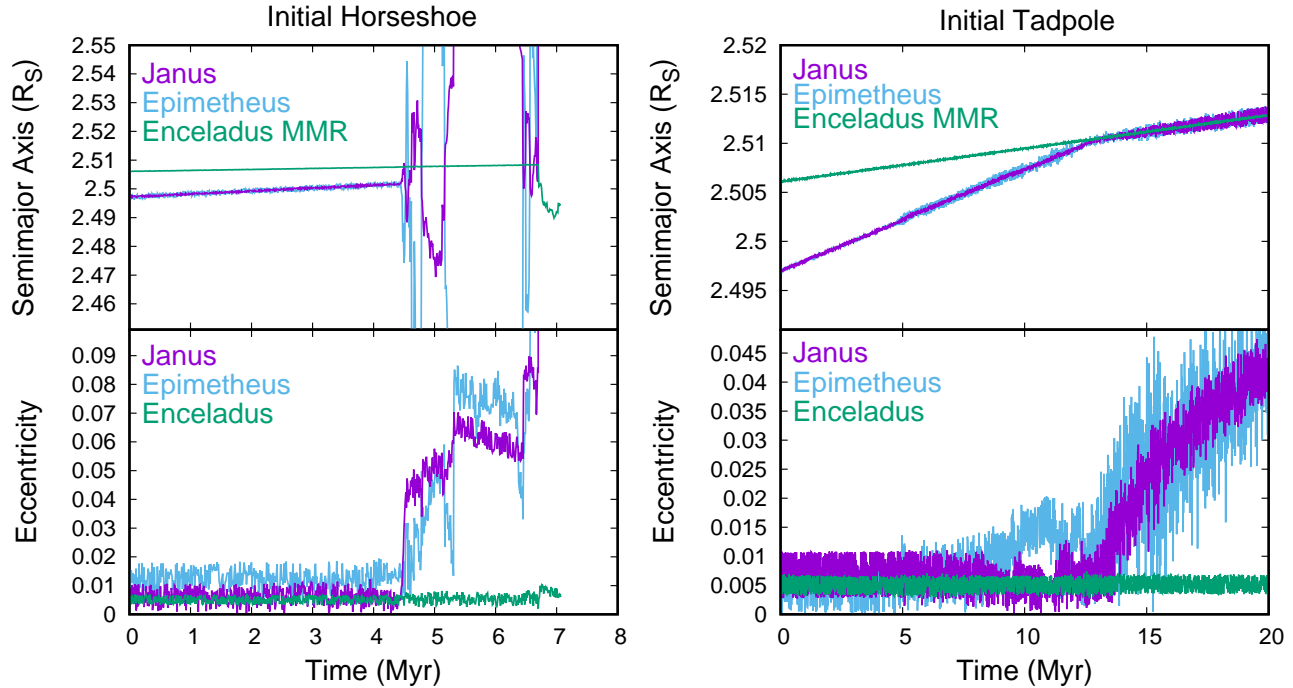}
\caption{Simulations of the 2:1 mean-motion resonance crossing between the coorbitals Janus and Epimetheus and Enceladus using {\sc simpl}. Enceladus and Dione were assumed to be on their current orbits, while Janus and Epimetheus were shifted inward. In the left-hand side panels Janus and Epimetheus were in their current horseshoe configuration, while they were put in a tadpole (Trojan) configuration in the right-hand panels. The coorbitals' orbits are expanding due to an artificial acceleration meant to represent ring torques. We used $k_2/Q=0.01$ for Enceladus, and we ignored satellite tides for the coorbitals. \label{janep}}
\end{figure*}

We simulated this resonance crossing between coorbitals and Enceladus using numerical integrator {\sc simpl}, previously used by \citet{cuk16}. Enceladus was assumed to be in its present orbital resonance with Dione. At first we assumed that Janus and Epimetheus were in a horseshoe configuration before the resonance, and gave them their present eccentricities and inclinations. Left-hand panels in Fig. \ref{janep} show a typical outcome of such simulations, in which the horseshoe configuration is disrupted and Janus and Epimetheus experience close encounters (our integrator does not test for collisions). This destabilization happens well before the core of the 2:1 resonance with Enceladus is reached, implying that the horseshoe is disrupted by near-resonant perturbations. 

Next we assumed that the two moons were corbitals before the resonance, but in a Trojan or ``tadpole'' configuration \citet{md99} (typical results shown in right-hand side panels of Fig. \ref{janep}). Interestingly, near-resonant perturbations always convert the tadpole configuration into a horseshoe, which is stable from there on. However, the moons are then captured into the 2:1 resonance with Enceladus, in which their eccentricities grew over time while the semimajor axis is locked to that of Enceladus. This eccentricity growth is largely unaffected by any tidal damping, as the small sizes and demonstrably rigid natures of irregularly shaped Janus and Epimetheus do not allow for significant tidal deformation or dissipation over $10^7$~year timescales we are interested in. We expect the resonance eventually to break and Janus and Epimetheus to collide. As we expect the moons to have large eccentricities at the time of the collision while their semimajor axes are initially within the resonance with Enceladus, the resulting debris must re-accrete significantly interior to the resonance (due to angular momentum conservation). 

\begin{figure*}
\epsscale{1.2}
\plotone{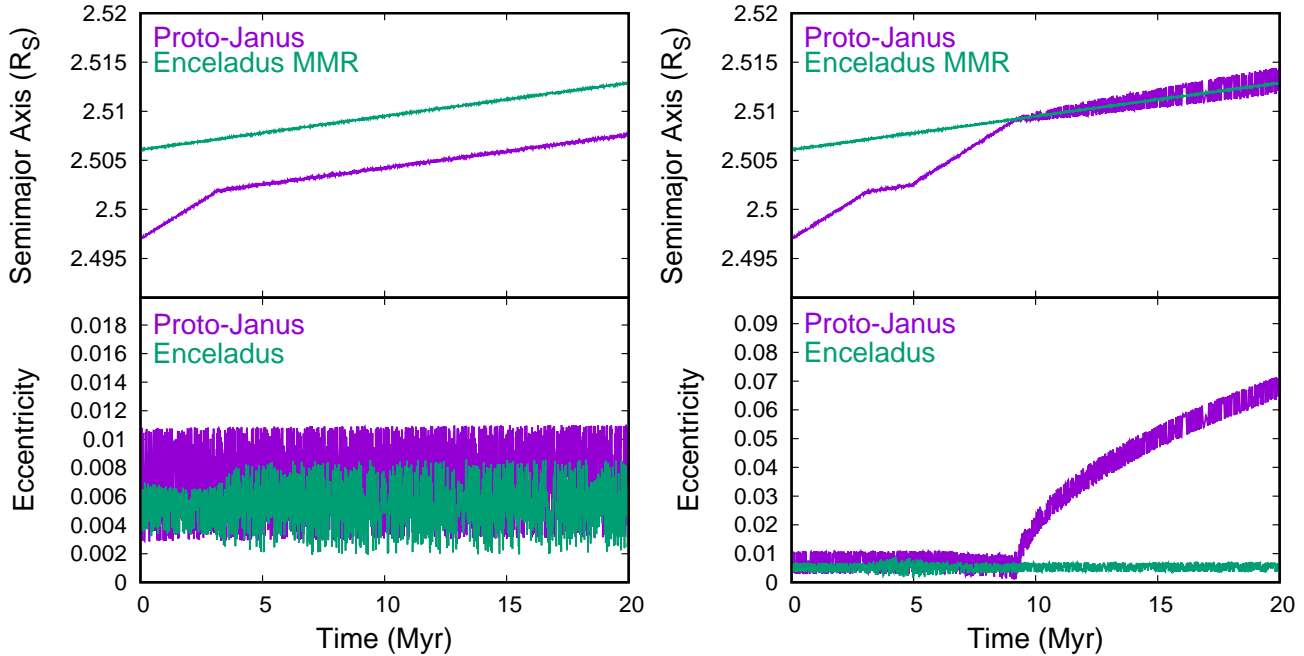}
\caption{Simulations of the 2:1 mean-motion resonance crossing between the parent body of the coorbitals Janus and Epimetheus and Enceladus using {\sc simpl}. The coorbitals' parent body was given their combined mass and was put on an orbit interior to Enceladus resonance with the current $e$ and $i$ of Janus. The progenitor's orbit is expanding due to an artificial acceleration approximating ring torques. The two simulations differ slightly in initial conditions. We used $k_2/Q=0.01$ for Enceladus, and we ignored satellite tides for the coorbitals' precursor.\label{protoje}}
\end{figure*}

We also ran a simulation assuming that Janus and Epimetheus were contained within one body during the resonance crossing with Enceladus (which may have subsequently broken up). Our simulations of the 2:1 resonance crossing with Enceladus show that the precursor is always captured in the resonance (Fig. \ref{protoje}). Note that we assumed current orbital eccentricity of Janus for the progenitor, and the present-day configuration of the Enceladus-Dione resonance. In most simulations, proto-Janus is captured into a stable corotation resonance in which the eccentricity of Enceladus is slightly higher than equilibrium eccentricity in the resonance with Dione (left-hand panels in Fig. \ref{protoje}). In some cases this resonance is broken and proto-Janus is captured into the $e_E$(i.e. Lindblad) sub-resonance in which the eccentricity of proto-Janus grows over time, like we saw for the initial tadpole configuration. Given the large orbital precession rates of the innermost moons, all resonances are well separated and there is no obvious way of breaking this lock purely through dynamics. However at eccentricities in the $0.05-0.1$ range proto-Janus would cross the orbits of Prometheus and Pandora (if they were present at that epoch) or the outer edge of the rings. Resulting collisions would presumably break the resonant lock between proto-Janus and Enceladus. Once again, angular momentum conservation dictates that any re-accretion of proto-Janus's debris happens interior to the resonance with Enceladus.   


If we assume that the coorbitals were produced in a breakup of a single progenitor, we can use their orbits to constrain the orbit of this progenitor. Using current orbits for the coorbitals, and under assumption of conservation of momentum at separation, we find that pre-breakup parent body of Janus and Enceladus had an eccentricity $e \leq 0.006$, much less than expected from the past capture into the 2:1 resonance with Enceladus. Eccentricity can be erased through re-accretion, but as we discuss above, disruption and reaccretion move the resulting new moons interior to the resonance with Enceladus. Therefore, either the tidal evolution of Enceladus has to be comparable with the orbital evolution of the horseshoe pair, or the inner Saturnian system has to be rather young (a combination of these two factors is also possible). 

If the resonant pair Enceladus and Dione are evolving through equilibrium tides with $Q/k_2=5000$ for Saturn \citep{lai12}, their joint orbital evolution timescale would be $a/\dot{a} \approx 10$~Gyr. Enceladus being locked to an inertial mode would produce the same timescale (even if the amount of tidal heating would be very different, as all of Dione's orbital evolution would now be due to Enceladus). This rate of evolution would only push the resonance with Janus-Epimetheus precursor back to 40 Myr. Locking of Enceladus to a Saturn-frame mode would lead to a $a/\dot{a} \simeq 130$~Gyr, which would produce negligible migration compared to the ring-torque-driven evolution of the coorbitals' precursor, keeping the apparent age of the resonant encounter at 30 Myr ago. Enceladus's rate of evolution through equilibrium tides (but not resonance locking) depends on its mean-motion resonance with Dione. Before the resonance Enceladus would have had an equilibrium tidal evolution timescale $a/\dot{a}=5$~Gyr (assuming $Q/k_2=5000$ for Saturn), so depending on the age of the Enceladus-Dione resonance, the progenitor-Enceladus resonance would have happened between 40 and 75~Myr ago. 

\citet{lai20} propose that a model in which equilibrium tides are weak and the moons' orbits evolve solely through resonance locking may result in a long-term stable system. In this model, the moons' orbits are not expected to converge and enter mean-motion resonances with each other. This way dynamical excitation and possible destabilization through MMRs is avoided. The simulations discussed here show that the inclusion of ring torques introduces relatively rapid convergence of the orbits of some of the inner moons. A young age ($<50$~Myr), at least for the Janus-Epimetheus parent body (JEPB) if not for other moons, appears inevitable. If we assume the presence of both equilibrium tides and resonant tidal response (Section \ref{ssec:bern}) the actual maximum age for JEPB may be somewhat older than 75~Myr but difficult to calculate precisely. Also, it appears that the JEPB formed close to its present location, and that the idea that Janus and Epimetheus evolved from the rings smoothly to their present position \citep{cri12} may not be tenable given the apparent impossibility of the pair (or their parent body) crossing the 2:1 resonance with Enceladus. 

\subsection{Establishment of the Current Mimas-Tethys 4:2 Resonance}\label{ssec:mt42}

Mimas and Tethys are currently in a 4:2 inclination-type resonance that involves the inclinations of both moons (the resonant argument is $4\lambda_{\Theta} - 2\lambda_M - \Omega_M - \Omega_{\Theta}$), where the subscripts $M$ and $\Theta$ refers to Mimas and Tethys, respectively. Large libration amplitude of the resonance has in the past been used to suggest that the resonance capture was a low-probability event \citep{all69, sin72}. Our work so far (under the assumption of equilibrium tides) suggests that there are at least two distinct dynamical pathways to resonance capture, both of which require later increase of the resonance argument libration width due to a third body. 

\begin{figure*}
\epsscale{1.2}
\plotone{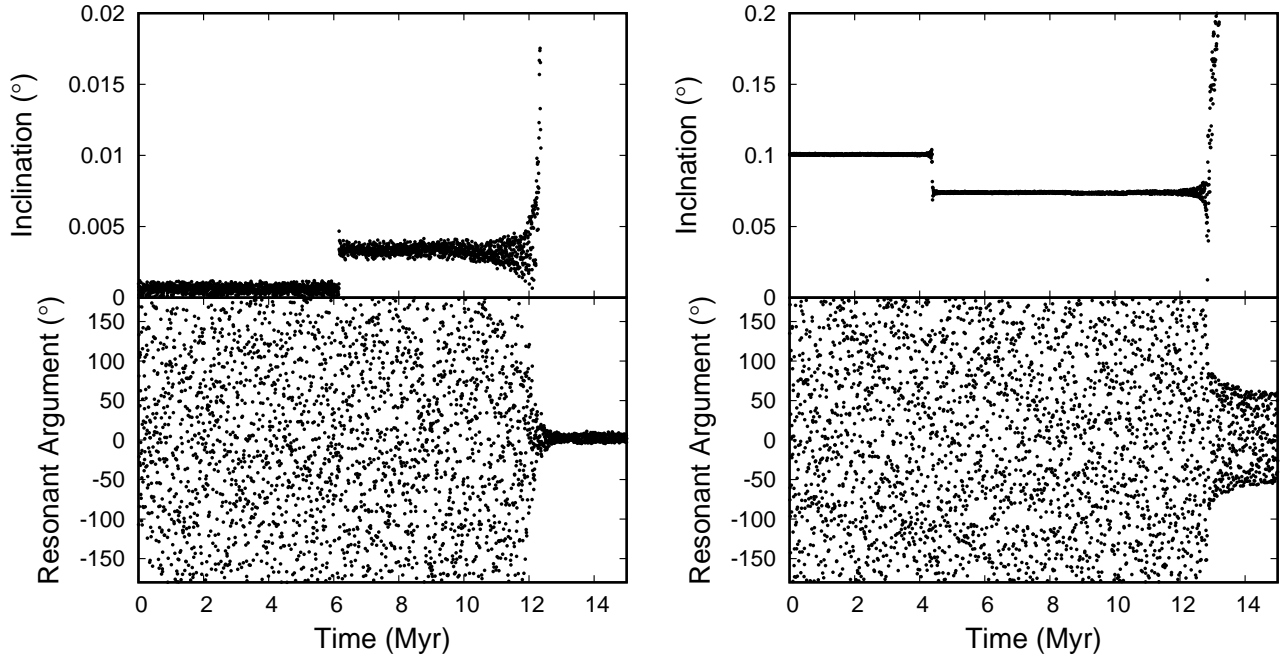}
\caption{Simulations of the 4:2 mean-motion resonance crossing and capture between Mimas and Tethys. The resonant argument plotted in bottom panels is $4\lambda_{\Theta} - 2\lambda_M - \Omega_M - \Omega_{\Theta}$. In the left-hand simulation Mimas was assumed to initially have a very low inclination, while in the right-hand simulation we assumed initial $i_M=0.1^{\circ}$ for Mimas. We assumed equilibrium tidal evolution with a frequency-independent $Q/k_2=4000$ for Saturn, $Q/k_2=10^5$ for Mimas and $Q/k_2=10^4$ for Tethys. The first jump in inclination is due to the $i^2_M$ resonance.
\label{mimteth}}
\end{figure*}

If the initial inclination of Mimas was very low ($i_M \simeq 0.001^{\circ}$), lack of capture into the preceding $i_M^2$ sub-resonance and capture into $i_M i_{\Theta}$ sub-resonance are highly likely events (assuming $i_{\Theta} \approx 1^{\circ}$ before the resonance). Right-hand panels in Figure \ref{mimteth} shows a typical capture into the Mimas-Tethys resonance assuming equilibrium tides with $Q/k_2=4000$ for Saturn. This rate of tidal evolution implies that the current Mimas-Tethys resonance is only 20-25~Myr old, consistent with a relatively young proposed age of the system \citep{cuk16}. The physical reason behind the lack of capture into the $i_M^2$ sub-resonance appears to be non-adiabatic nature of the resonance crossing, as proposed by \citep{sin72}. While the tidal evolution is smooth and slow by most criteria, the extreme narrow width of the $i_M^2$ sub-resonance at very low inclinations seems to allow for non-resonant crossing.

Alternatively, if we assume pre-resonant inclination of $i_M=0.1^{\circ}$ for Mimas, there is a high probability (about $70\%-80\%$) of missing capture into the $i_M^2$ sub-resonance, and a similarly high probability of capture into the $i_M i_{\Theta}$ sub-resonance. In this case the resulting libration amplitude of the resonant argument is about 60$^{\circ}$, still short of the observed one but much larger than in the low inclination limit. In this regime the capture into $i_M^2$ is unlikely because of large initial inclination \citep[as is common for resonant encounters][]{md99}, while the capture into $i_M i_{\Theta}$ sub-resonance is somewhat more likely due to large $i_{\Theta}$ (also, negative kick to the $i_M$ when crossing $i_M^2$ also makes the next capture more likely). 

\begin{figure*}
\epsscale{.8}
\plotone{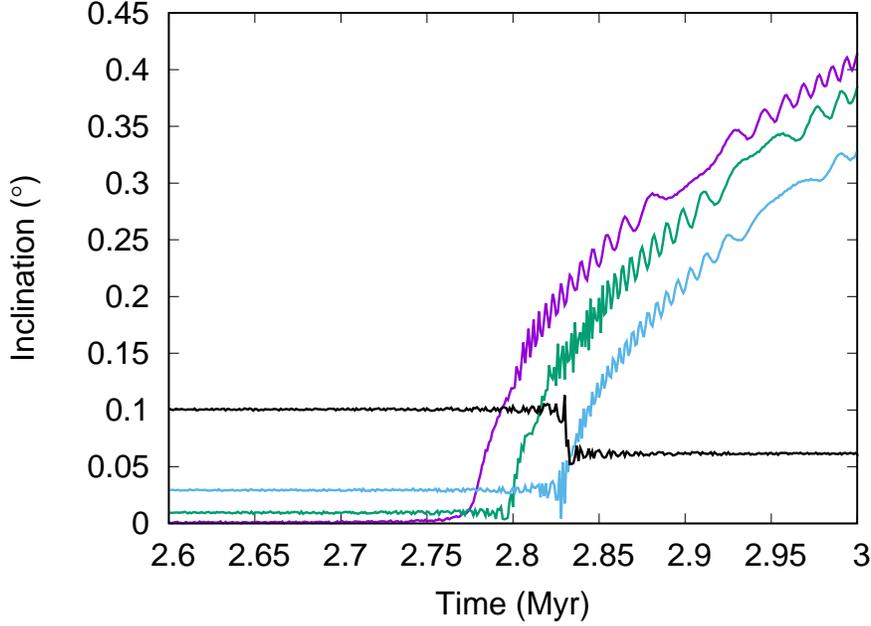}
\caption{Simulations of the crossing and capture into a third order sub-resonance of the Mimas-Tethys 2:1 MMR. The inclination of Mimas is plotted in four different simulations which started with $e_{\Theta}=0.002$ and $i_{\Theta}=1^{\circ}$ for Tethys, and $e_M=0.02$ and a range of initial inclinations for Mimas. The argument of this resonance is $2\lambda_{\Theta}-\lambda_M+\varpi_M-\Omega_M-\Omega_{\Theta}$. Satellite $k_2$ and $Q$ same as Fig. \ref{mimteth}.\label{3incres}}
\end{figure*}

Figure \ref{mimteth} shows two of the many simulations we did to study capture into the 4:2 Mimas-Tethys resonance in isolation, meaning that we did not simulate the system's evolution before and after this encounter. Initially we expected the two sub-resonances of the 4:2 MMR ($i^2_M$ and $i_M i_{\Theta}$) to be the only relevant inclination-type terms of the 2:1 Mimas-Tethys commensurability. Later simulations of the comprehensive recent dynamical history of the system have revealed additional relevant inclination-type resonances. In particular, a third-order harmonic of the 2:1 Mimas-Tethys resonance with the argument $2\lambda_{\Theta}-\lambda_M+\varpi_M-\Omega_M-\Omega_{\Theta}$ is surprisingly strong, and can lead to capture for low initial inclinations of Mimas (see Fig. \ref{3incres}). This sub-resonance (which we designate $e_M i_M i_{\Theta}$) is encountered well before the $i_M^2$ and $i_M i_{\Theta}$ sub-resonances shown in Fig. \ref{mimteth}. The reason for the unexpected strength of $e_M i_M i_{\Theta}$ resonant term are the high inclination of Tethys and eccentricity of Mimas, which we assumed both predate these moons' 4:2 MMR. These assumptions were based on lack of available mechanisms of exciting $e_M$ after the capture into the current 4:2 Mimas-Tethys resonance, and the necessity of large $i_{\Theta}$ for the capture to occur. We find that an initial $i_M \approx 0.1^{\circ}$ is necessary in order to avoid capture into the $e_M i_M i_{\Theta}$ sub-resonance. Therefore, while both low- and high-inclination routes to capture into the observed $i_M i_{\Theta}$ resonance are possible in isolation, additional constraints from the $e_M i_M i_{\Theta}$ crossing make the high inclination path the only likely capture mechanism. 

Captures into resonance shown in Fig. \ref{mimteth} results in a libration amplitude of the resonant argument that is smaller than observed. In general, we find that that the libration amplitude of the Mimas-Tethys resonance is highly vulnerable to perturbations from other moons (including Enceladus, Sec. \ref{ssec:et118}), so this resonance's libration amplitude is less indicative of capture mechanism than previously thought. We note that the difference between the post-capture libration amplitude and the observed value is less pronounced in our preferred high inclination route to Mimas-Tethys 4:2 MMR compared to the low-inclination route. Clearly more detailed study of the system's very recent evolution, accounting for the influence of all the moons (including Tethys's small Trojan companions), is necessary to better understand the evolution of the libration amplitude of the Mimas-Tethys resonance. 


All hypotheses of the origin of the Mimas-Tethys resonance must include a convergent evolution of these two moons. This convergent evolution is expected in the case of equilibrium tides, but interestingly only in the ``constant Q'' rather than ``constant time-lag" model (as the synodic period of Mimas is relatively long). In case of evolution by both moons being locked to inertial waves there should in principle be no relative orbital evolution, either convergent or divergent, while resonance locking to Saturn-frame modes would result in divergent evolution. If only Mimas is in resonant lock, then a resonance would form and should be about 20 Myr old (in the case of inertial waves) or about 300 Myr old (in case of locking to Saturn-frame modes). Locking of Mimas to a Saturn-frame mode would require a ``background'' $Q \ge 100,000$, as faster equilibrium tidal migration would allow Mimas to ``outrun'' the mode.

Another common assumption in all scenarios on the origin of the Mimas-Tethys resonance require Tethys to have a prior inclination of about $i_{\Theta}\approx 1^{\circ}$. This relatively large inclination requires some kind of past dynamical interaction with other moons, most likely a resonance. \citet{cuk16} have found this inclination to be a plausible consequence of Dione-Rhea 5:3 resonance crossing, followed by the (associated) Tethys-Dione secular resonance. We note that this scenario requires convergent orbital evolution of Dione and Rhea, at least in the past (see Subsection \ref{ssec:rhea})

\subsection{Observed Acceleration of Rhea and the Past 5:3 Dione-Rhea Resonance}\label{ssec:rhea}

The orbit of Rhea is directly observed to evolve on 6 Gyr timescale \citep{lai17,lai20, jac22}. The rate of orbital evolution of Titan may be similarly fast \citep{lai20}, but this is still disputed \citep{jac22}. Observational results for Rhea are clearly not consistent with the expectations based on equilibrium tides. Locking to resonant modes within Saturn has been proposed as an explanation for these observations \citep{ful16}, with the apparent rate of the moon's orbital evolution being determined by the drift in frequency of the resonant peaks in the tidal response of the planet. After Rhea's fast evolution was first observed \citep{lai17}, the assumption that the peaks in tidal response have a constant absolute frequency drift in Saturn's rotating frame \citep{ful16} implied that Titan's own drift timescale should be $a/\dot{a} < 2$~Gyr. The observed tidal evolution rate of Titan of 11~Gyr \citep{lai20} would imply that the moons could not both be in resonant lock, or that the resonant mode drift is not constant in the rotating frame. Locking to inertial waves which drift together in an inertial reference frame was proposed instead \citep{lai20}, which should result in the same tidal evolution timescale for all moons in resonant lock. The difference between the apparent evolution timescales for Rhea and Titan reported by \citet{lai20} seems to suggest that moons under resonant lock can still have converging orbits, which would challenge the big picture of \citet{lai20}, in which moons never cross mutual resonances. On the other hand, if Titan has a much slower orbital evolution consistent with equilibrium tides \citep[as found by][]{jac22}, then we can be sure that not all moons are locked to resonant modes. Clearly the dynamics of the system is more complex than predicted in any one simple model.

The observed fast evolution of Rhea is at odds with the hypothesis that Dione and Rhea crossed their mutual 5:3 MMR. The Dione-Rhea 5:3 MMR crossing was proposed by \citet{cuk16} as a way of producing the current inclinations of Tethys and Rhea. The inclination of Tethys is particularly significant ($i_{\Theta} \approx 1^{\circ}$) and must have predated Tethys's current resonance with Mimas. As \citet{cuk16} show, the Dione-Rhea 5:3 resonance is usually followed by a secular resonance between Tethys and Dione during which Dione ``passes'' its eccentricity and inclination to Tethys. Therefore there is a compelling reason to think that a past Dione-Rhea 5:3 MMR crossing did take place.

The orbital evolution of Rhea therefore presents a conundrum. Its rate is much too fast for equilibrium tides (assuming $Q/k_2=5000$), but it is also too fast for resonance locking \citep[assuming that Titan is currently in resonance lock][]{lai20}. A Saturn-frame mode resonant lock would have $a/\dot{a} \approx 40$~Gyr, while resonance lock with inertial modes would predict $a/\dot{a}=11$~Gyr, same as Titan. Furthermore, the apparent past crossing of the Dione-Rhea resonance implies that this fast evolution of Rhea is a very recent phenomenon (i.e. over the past several Myr so so). Assuming all of these constraints are correct, the simple solution is that Rhea is not locked to a mode, but is currently {\it crossing} a resonant mode. This is a possible situation if a moon (through equilibrium tides) evolves faster than a mode rather than the other way around, making a resonant lock impossible but producing occasional bursts of fast orbital evolution. If we assume that Titan is currently evolving with a timescale of 100~Gyr \citep[consistent with equilibrium tides][]{jac22}, then it is possible that resonance lock may not be feasible at all. The argument here is that Titan would have certainly been captured in a resonant lock if resonant modes move faster than Titan's orbit evolves due to equilibrium tides. In principle, we can also envision resonant modes moving to higher frequencies over time, opposite the direction of tidal evolution. In this case, all accelerated tidal evolution would always be episodic, being driven by the moons overtaking resonant modes. 

In the rest of the paper, we will examine how a combination of equilibrium tides and passage through resonant modes can explain other features of the Saturnian system, namely the establishment of the Enceladus-Dione MMR. 



\section{Enceladus-Dione 2:1 Resonance: Equilibrium Tides}\label{sec:equil}

In the previous sections we concluded that orbital evolution through equilibrium tides (with $Q/k_2=5000$) is able to explain the present heating of Enceladus \citep{mey07} and the capture of Mimas and Tethys into their current resonance with a high probability, given appropriate initial conditions. We also find that orbital evolution of Enceladus through equilibrium tides allows for an age of Janus-Epimetheus pair (or their parent body) longer than 40~Myr, unlike Enceladus's evolution through resonant lock. However, it has not been explored in the prior literature whether equilibrium tides can establish the current Enceladus-Dione 2:1 MMR, as opposed to maintaining it. Study of this resonant capture must be done through numerical integrations, as \citet{cuk22} have shown that practically all two-body resonances in the Saturnian system harbor three-body resonances that are very hard to model analytically, in addition to many two-body sub-resonances. 

\subsection{Enceladus-Dione 4:2 Inclination Resonances}\label{ssec:ed42}

\begin{figure*}
\epsscale{1.2}
\plotone{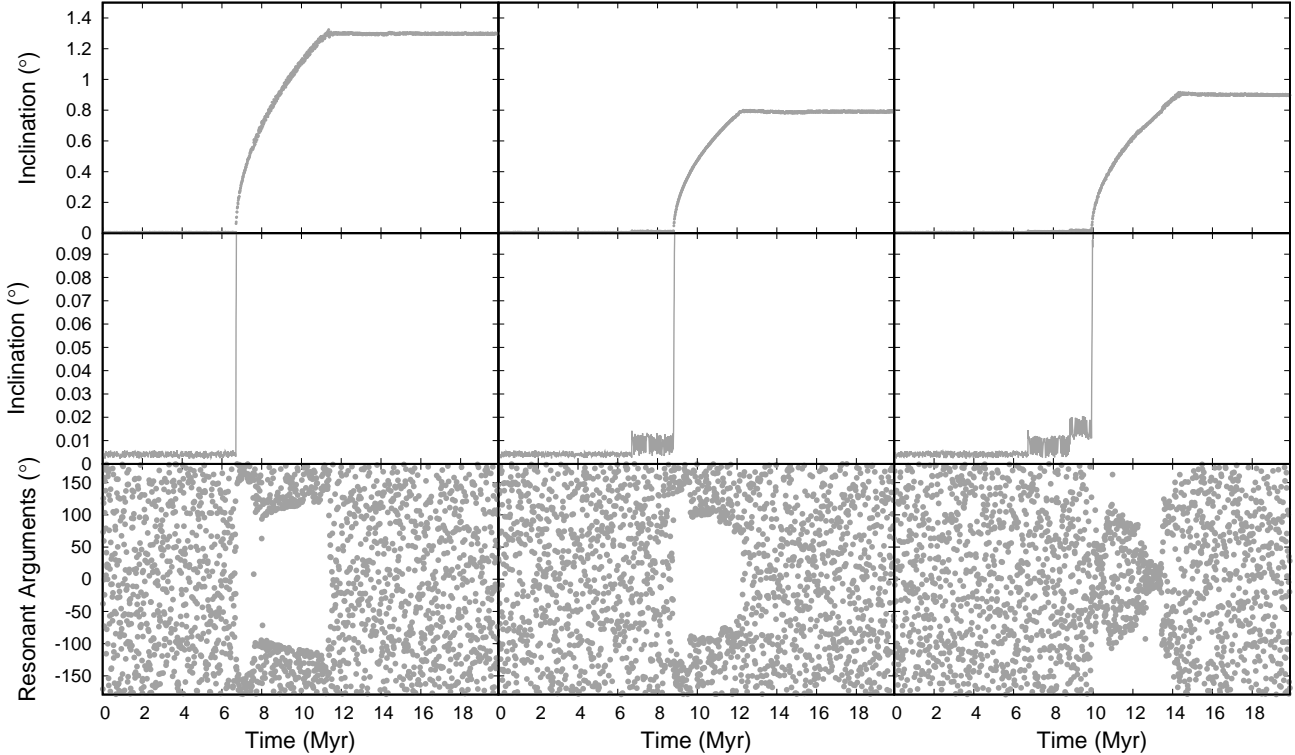}
\caption{Simulations of the 4:2 mean-motion resonance crossing and capture between Enceladus and Dione. Left hand panels show capture into the pure Enceladus inclination resonance $i_E^2$, middle panels in Fig. \ref{seliv} show capture into the semi-secular three-body resonance involving Tethys with an argument $4\lambda_D-2\lambda_E-\Omega_E-\Omega_{\Theta}$, and the right-hand panels show capture into the mixed two-body 4:2 $i_E i_D$ sub-resonance. We used $Q/k_2=10^4$ for all three moons. \label{seliv}}
\end{figure*}

The Enceladus-Dione 2:1 MMR contains two first-order sub-resonances, four more second-order ones, several important three-body resonances, as well as numerous two-body third-order sub-resonances. The currently occupied $e_E$ sub-resonance is one of the last to be encountered as the orbits of Enceladus and Dione converge. We ran a number of simulations of the Enceladus-Dione 2:1 MMR encounter assuming equilibrium tides and $Q/k_2=4000$ for Saturn. Typically, Enceladus was captured into the 4:2 $i_E^2$ resonances in large majority of cases where Enceladus and Dione encounter this resonance under equilibrium tides. Left-hand panels in Fig. \ref{seliv} plot the result of a typical simulation that features capture into the Enceladus-Dione 4:2 $i_E^2$ resonance. Enceladus's inclination grow to about $1^{\circ}$, at which point secondary resonances break the inclination resonance. In this particular simulation all inner moons (from Mimas to Rhea) were assumed to have initially equatorial orbits, but the result is the same when Tethys was assumed to have been already inclined. In all remaining simulations of equilibrium tide Enceladus-Dione 4:2 MMR crossing Enceladus is caught into a more complex inclination resonance, also affecting the inclination of Dione or Tethys. In all cases Enceladus acquires an inclination on the order of a degree, two orders of magnitude above the observed one.

Near-certainty of capture into one of the inclination sub-resonances found in our simulations strongly suggests that the Enceladus-Dione 2:1 MMR was not assembled through equilibrium tides. This is in contrast to what we found for the Mimas-Tethys 4:2 resonance (Section \ref{mimteth}) which can be accounted for using equilibrium tides (except for the present libration amplitude, which is easily excited by a third body). High probability of capture into the Enceladus-Dione 4:2 $i_E^2$ resonance, but not Mimas-Tethys 4:2 $i_M^2$ resonance is rather surprising but holds in large numbers of numerical simulations. We suspect that the lack of capture into the Mimas-Tethys 4:2 $i_M^2$ resonance with low initial $i_M$ is due to the mechanism originally proposed by \citet{sin72}, which states that in cases of small free inclination and fast orbital precession even relatively slow resonant encounter may not be adiabatic, i.e. a second-order resonance may not be able to dominate the motion of the ascending node against oblateness-driven precession. Note that this explanation did not hold in numerical simulations using $Q \approx 18,000$ for Saturn \citep{sin74}, but appears to explain the behavior at ten times faster tidal evolution used in this study.

\subsection{Enceladus-Tethys 11:8 Resonance}\label{ssec:et118}

\begin{figure*}
\epsscale{1.}
\plotone{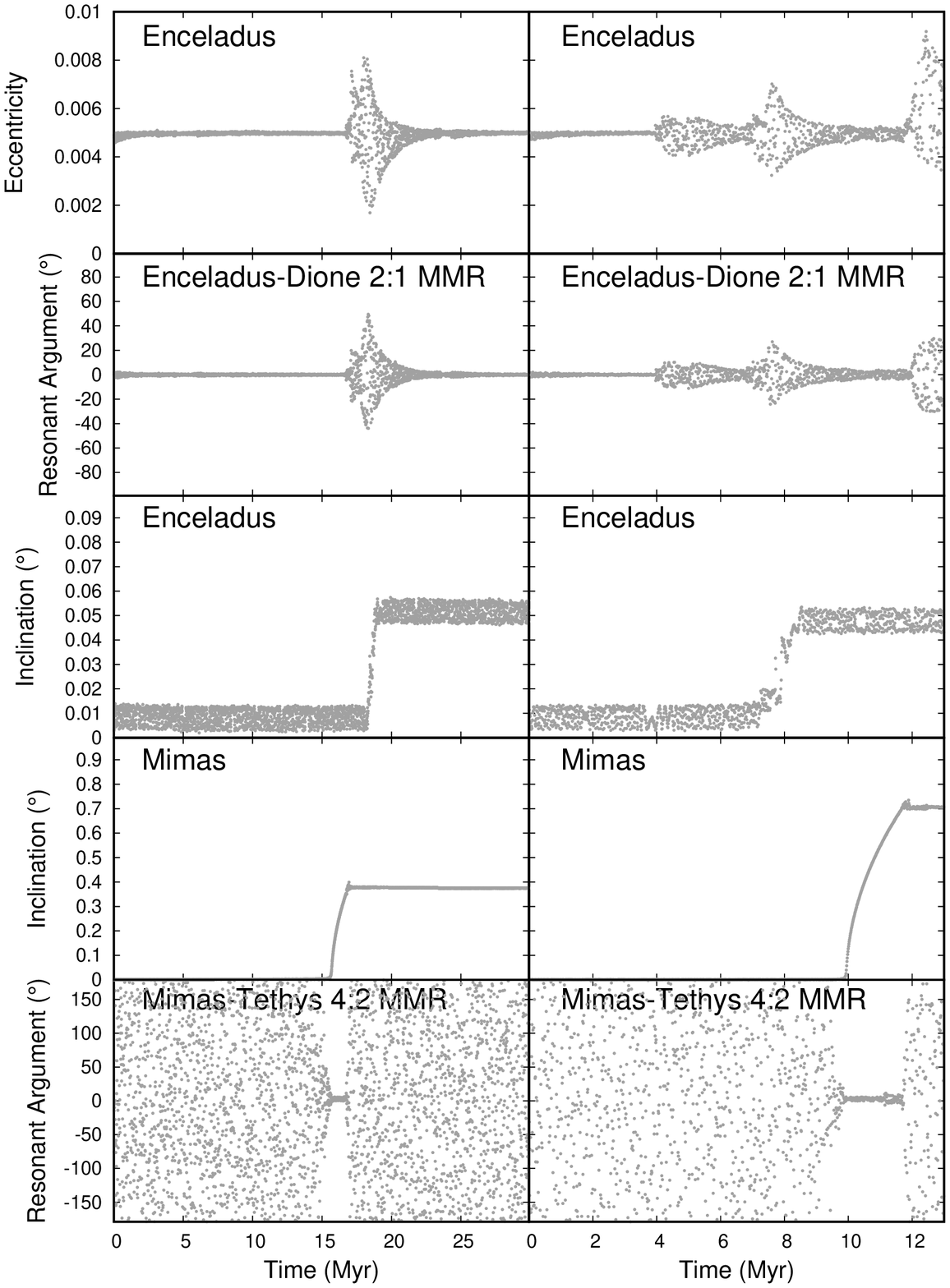}
\caption{Left-Hand side: A simulation of recent evolution of he inner moons assuming frequency-independent equilibrium tides with $Q/k_2=4000$ for Saturn. Initial conditions are set at about 30~Myr ago, and we assume that Enceladus-Dione 2:1 MMR ($2\lambda_D-\lambda_E-\varpi_E$) was already established and in equilibrium. Mimas and Tethys are captured into their 4:2 MMR ($4\lambda_{\Theta}-2\lambda_M-\Omega_M-\Omega_{\Theta}$) at 15~Myr, but this resonance is prematurely broken by the interaction through the 11:8 Enceladus-Tethys MMR, while Enceladus acquires significant inclination. Right hand side: A very similar simulation assuming that tidal dissipation at Tethys's frequency is only 70\% of that for other moons. In this simulation Enceladus suffers inclination excitation before the beginning of the Mimas-Tethys 4:2 MMR, confirming that the (already inclined) Tethys is the relevant perturber, rather than (still non-inclined) Mimas. We used $Q/k_2=10^5$ for Mimas, $Q/k_2=10^2$ for Enceladus, and $Q/k_2=10^4$ for Tethys and Dione.
\label{11to8}}
\end{figure*}

In the course of our work on the past dynamics of the Saturnian system, we discovered that Enceladus-Tethys 11:8 resonance offers unexpected constraints on the past orbital evolution of those two moons. In the equilibrium tide paradigm, this resonance would have been crossed 10-15 Myr ago. Since Tethys's predicted equilibrium tidal evolution is faster than that of Enceladus (they are the only neighboring pair of major Saturnian moons on diverging orbits, as Tethys's six times larger mass boosts its tidal evolution), this resonance should have been crossed divergently without a chance of capture. Despite the third order of this sub-resonance, the large eccentricity of Enceladus and inclination of Tethys make this resonant term surprisingly relevant. In particular, we find that the Enceladus-Tethys 11:8 MMR crossing usually both excited the inclination of Enceladus beyond the observed value, and disrupted the Mimas-Tethys 4:2 MMR (Fig. \ref{11to8}).  

These results decisively argue that the current system could not have resulted by evolution through equilibrium tides. Of course, observations of the tidal acceleration on Rhea \citep{lai17, lai20, jac22} already established that Saturn's tidal response is dynamic, and that Rhea and/or Titan may be in resonance lock. However, as we demonstrated in previous sections, the observed resonances between the inner moons appear to suggest convergent tidal evolution consistent with equilibrium tides. The apparent lack of past crossing of the Enceladus-Tethys 11:8 MMR is the first evidence from mutual resonances that requires frequency-dependent tides to have acted in the past.

In general, the Enceladus-Tethys 11:8 MMR could have been avoided through slower orbital evolution of Tethys or faster migration of Enceladus. We explored the former possibility assuming that the tidal response of Saturn is frequency-dependent but does not contain resonant modes. This would be consistent with findings of \citet{lai20} and \citet{jac22}, who do find slower orbital evolution of Tethys relative to other inner moons than expected from equilibrium tides (however, the uncertainties are still large). The current tidal evolution rate of Enceladus is constrained by its tidal heating (Sec. \ref{sec:heat}), so a slower evolution of Tethys is necessary to avoid a recent passage through the Enceladus-Tethys 11:8 MMR. However, we have just seen that the equilibrium tides acting on Enceladus are still unable to reproduce the Enceladus-Dione 2:1 MMR (Subsection \ref{ssec:ed42}). Therefore it appears that the most likely solutions to the problem posed by Enceladus-Tethys 11:8 MMR requires 1) both that the tidal response at Tethys's frequency is currently weaker than that of Enceladus, and 2) that the migration of Enceladus was much faster in the past. We explore this possibility in the next Section.

\section{Enceladus-Dione 2:1 Resonance: Short-Lived Enhanced Tidal Evolution}\label{sec:encmode}

In this section we present simulations of Enceladus and Dione encountering their 2:1 MMR while Enceladus is migrating much faster than expected by equilibrium tides. This is enabled in the numerical integrator {\sc simpl} by adding a simple factor multiplying the tidal acceleration of each moon. For moons evolving under equilibrium tides this factor is simply 1, and it is larger for accelerated evolution. We can use this modification of equilibrium tides as our hypothesis is that Enceladus (and later Rhea) are simply passing through frequency bands that have very high tidal response, rather than being locked to these resonant modes. Resonant locking cannot be modeled this way, as moons locked to resonant modes tend to have constant migration rate but do not experience constant tidal $Q$ of Saturn (and the apparent tidal $Q$ will change if the moon in question enters a resonance with an exterior satellite). 

\subsection{Inclination-Type Sub-Resonances of the Enceladus-Dione 4:2 MMR}\label{fast_ed42}

\begin{figure*}
\epsscale{1.}
\plotone{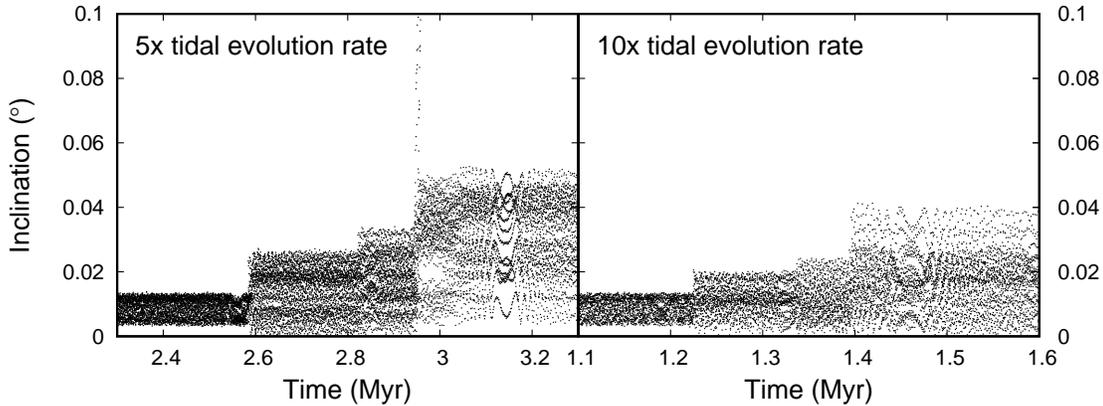}
\caption{Left-Hand side: Results of twenty simulations (plotted together) of the crossing of the inclination subresonances of the Enceladus-Dione 4:2 MMR with Enceladus's orbit evolving at 5x tidal rate. Right hand side: Twenty simulations of the same MMR crossing, now assuming Enceladus migrating at ten times the tidal rate. All the moons had their current inclinations and eccentricities at the start of the simulation, and each simulations had a different argument of pericenter of Enceladus, which led to spread in initial resonant arguments. Satellite tidal parameters were the same as in Fig. \ref{seliv}.\label{etv_inc}}
\end{figure*}

It is a well-known phenomenon that the probability of mean-motion resonance capture declines as the resonance is encountered more rapidly \citep{md99}. In this sub-section we explore the hypothesis that an episode of fast tidal evolution may have enabled Enceladus to cross the inclination resonances with Dione without capture. Results from a number of simulations are plotted in Fig. \ref{etv_inc}. We ran twenty different simulations that had Saturn's $Q/k_2=4000$ for all moons except Enceladus, for which the response was enhanced 5 times; six resulted in capture into an inclination-type resonance and fourteen experienced a kick in inclination. In twenty simulations where Enceladus evolved ten times faster than expected, none experienced capture. Post-passage inclination of Enceladus was on average lower in 10-times accelerated simulations (typically $i_E=0.01-0.02^{\circ}$) than in 5-times accelerated ones (typically $i_E=0.02-0.04^{\circ}$).

Our choice of current inclinations and eccentricities for initial conditions is somewhat arbitrary. While its present inclination is small, Enceladus presumably formed with close to zero inclination and this initial value may affect the post-resonant distribution of inclinations. Later simulations (shown in Section \ref{ssec:bern}) were started with a lower inclination of Enceladus but the outcomes were very similar. As seen in Fig. \ref{etv_inc}, the three-body resonance with Tethys (middle of the three jumps) is a minor contributor to Enceladus's final inclination, so our assumption that Tethys was already inclined at this epoch is not crucial. The exact inclination of Dione may effect the results somewhat (as the third jump is the $i_E i_D$ sub-resonance), but we see no strong reason to expect a different inclination for Dione at the time when 2:1 resonance with Enceladus was established. 

\subsection{Eccentricity-Type Sub-Resonances of the Enceladus-Dione 4:2 MMR}\label{fast_ed21}

After the inclination-type sub-resonances have been crossed, Enceladus encounters a number of eccentricity type sub-resonances of the Eneceladus-Dione 2:1 MMR. Based on results shown in Fig. \ref{etv_inc}, we enhanced Enceladus's tidal evolution by a factor of ten, while the tidal evolution of other moons corresponds to frequency-independent $Q/k_2=4000$ for Saturn. The choice of initial eccentricities requires making assumptions not only about past resonances but also about the rate of eccentricity damping by the moons. 

\begin{figure*}
\epsscale{1.}
\plotone{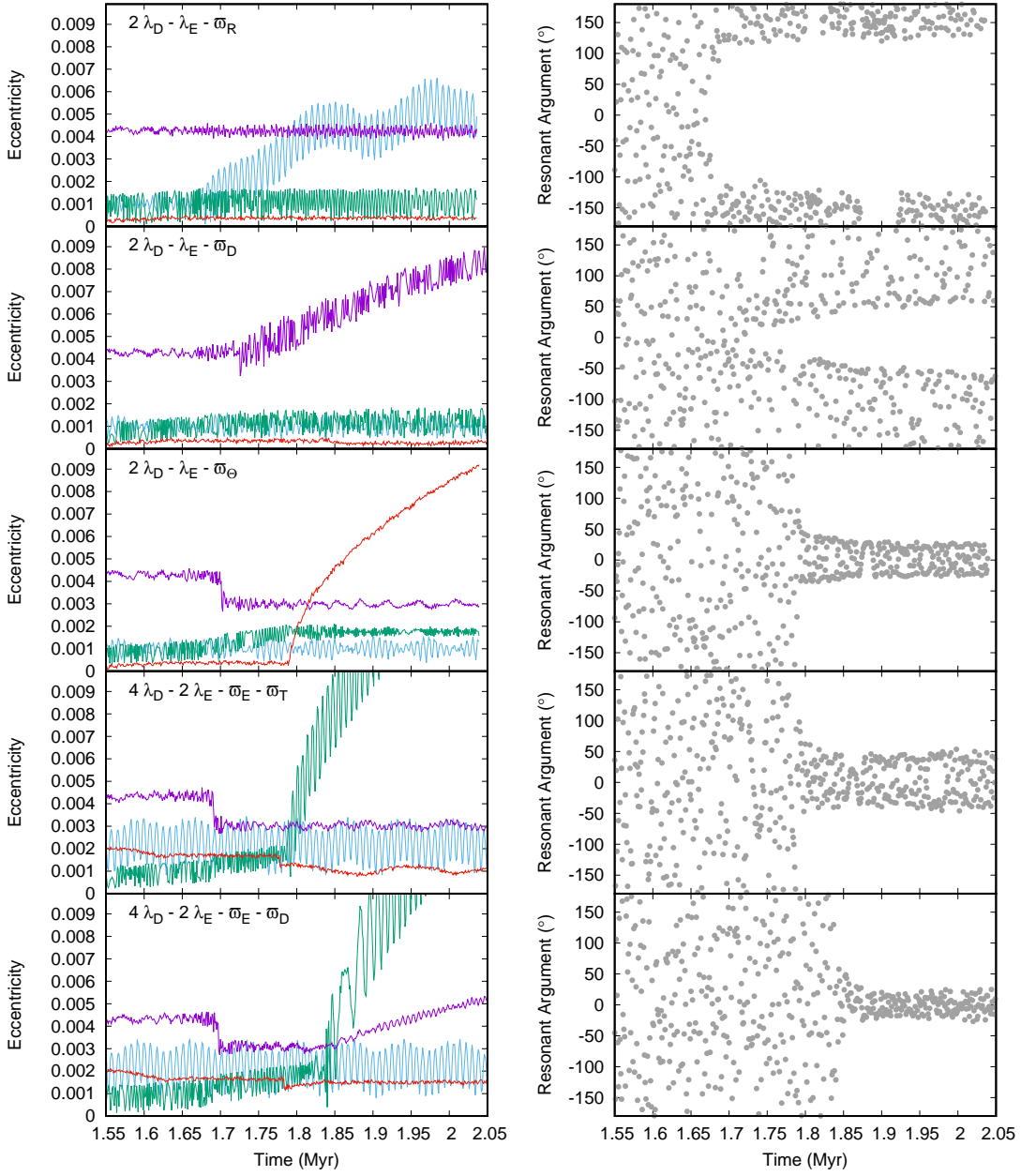}
\caption{Example simulations of captures into various eccentricity sub-resonances of the Enceladus-Dione 2:1 MMR. Left hand panels show the eccentricities of relevant moons: Enceladus (dark green), Tethys (red), Dione (magenta) and Rhea (cyan), while the right-hand panels plot resonant arguments which are labelled on the corresponding left-hand plots. In the top three rows we assumed that Tethys and Rhea had their current low eccentricities, and in the bottom two rows we assumed their eccentricities were higher. Enceladus's migration rate was 10x tidal, and its initial eccentricity was assumed to be very low. We conclude that low eccentricities of Tethys and Rhea lead to capture into first-order three-body resonances, while low $e$ of Enceladus leads to captured into second-order resonances (two- and three- body). These simulations are a subset of those shown in right-hand side of Fig. \ref{etv_inc}. 
\label{etv_ecc}}
\end{figure*}

Our first batch of simulations were using the current eccentricities for most moons, except that Enceladus was given a low initial $e_E<0.001$ while Dione was given $e_D=0.004$, about twice the current value (to account for the expected resonance kick and subsequent damping). Out of the first group of simulations using these initial conditions, the majority of runs resulted in Enceladus being captured into three-body resonances with Rhea or Tethys (Fig. \ref{etv_ecc}, first and third rows), and the rest resulted in $e_D$ sub-resonance, with no simulations reaching the current $e_E$ sub-resonance (which is last in order of encounters).

Captures into three-body resonances, which were understandably missed in past semi-analytical models \citep{mey08} demonstrate once again the importance of such resonances in the Saturnian system \citep{cuk22}. These captures may come as a surprise given the high rate of orbital migration of Enceladus, but we should recognize that these are first-order resonances, unlike the inclination-type resonance shown in Fig. \ref{etv_inc}, which are necessarily of the second order. This preponderance of three-body resonance captures is in part a consequence of us starting the simulations with Tethys and Rhea on very low eccentricity orbits. However, these were not arbitrary choices, as these two moons currently do have almost circular orbits. Low current eccentricities of both Tethys and Rhea argue against these three-body resonance captures happening in the recent past, as damping the eccentricities acquired in Fig. \ref{etv_ecc} in realistic timescales may be difficult, especially for Rhea. We therefore prefer these moons having relatively small eccentricities at this epoch and avoiding capture into the three-body resonances, rather than being captured and acquiring large eccentricities (on the order of 0.01 or more). 

In the second batch of simulations, we gave both Tethys and Rhea eccentricities of 0.002, which we found (through trial and error) to be largely sufficient to avoid capture into three-body resonances. In the second group of simulations, Enceladus still never reaches the current $e_E$ resonance, but in majority of cases becomes captured in a second-order resonance which also includes eccentricity of either Titan or Dione (Fig. \ref{etv_ecc}, rows four and five, respectively). While the mixed resonance $e_Ee_D$ has been studied before \citep{mey08}, the three-body resonance resonance with Titan is new. In both of these cases we cannot avoid capture into these sub-resonances by changing the eccentricity of Dione or Rhea. A less eccentric Dione would make capture into $e_E e_D$ resonance less likely, but would increase the fraction of captures into the first order $e_D$ sub-resonance (this fraction is already 20-30\% for initial $e_D=0.004$) and may conflict with Dione's current eccentricity. In the case of Titan, its eccentricity does not get affected by the resonance and we must use values very close to the current one.

The simpest way to avoid the capture into the above mentioned second-order eccentricity sub-resonances is to give Enceladus a sizable initial eccentricity. The implication is that at the time of capture into the Enceladus-Dione 2:1 MMR, not only there had to be an accelerated orbital evolution of Enceladus (to avoid capture into the inclination-type resonances), but all inner moons had to have somewhat eccentric orbits (to avoid capture into the eccentricity-type resonances). In the next section we will put together a possible evolution path that satisfies the constraints presented in this and previous sections, and the implications of the moons previously excited eccentricities will be discussed in Section \ref{discon}.

\subsection{Comprehensive Model of the System's Recent Evolution}\label{ssec:bern}

In this section we attempt to construct a numerical model of the recent (last 20~Myr) evolution of the Saturnian satellite system. We are using all of the constraints that were established in previous sections. Enceladus and Dione are assumed to experience tidal acceleration with $Q/k_2=4000$ at the present epoch,\footnote{we assumed a value on the high-end of this estimate of dissipation in order to make simulations faster.} in agreement with the hypothesis of equilibrium tidal heating of Enceladus \citep[Section \ref{sec:heat},][]{mey07, how11, lai12}. Mimas is assumed to have the same tidal evolution rate, both for lack of other constraints, and as it is consistent with Mimas encountering and being captured into 4:2 resonance with Tethys (Section \ref{ssec:mt42}). Need for avoidance of the 11:8 Enceladus-Tethys resonance, as well as (admittedly uncertain) results by \citet{lai20} and \citet{jac22}, suggested that the tidal acceleration of Tethys is much weaker than that expected from equilibrium tides. Finally, most radically, we assumed that Enceladus went through a stage of very rapid tidal evolution due to crossing of (but not locking to) a resonant mode. This was done to avoid capture into Enceladus-Dione 4:2 inclination resonances (Section \ref{ssec:ed42}), as well as to avoid a recent crossing of Enceladus's 1:2 resonance with the horseshoe moons (or their progenitor; Sec \ref{ssec:janus}). Additional assumptions include low initial tidal dissipation within Enceladus until it experiences strong tidal heating, at which point Enceladus switches to a strong tidal response (this is done to approximate melting of the interior). Due to the nature of the integrator where tidal response is treated as a constant parameter, we modelled both the transitions in the tidal acceleration of Enceladus and the changes in its own tidal response as abrupt events, implemented manually. 

Table \ref{initial} shows the initial conditions for two sets of simulations of the integrated evolution of the inner moons (sets differed only by initial $i_{\Theta}$). In both cases, Enceladus was assumed to be experiencing 10x accelerated evolution in the first 10~Myr of the simulation. After 3~Myr we selected one simulation in each set in which Enceladus was captured into the 2:1 $e_E$ MMR with Dione and acquired high eccentricity, cloned those simulations, and from that time onward used a much higher tidal response of Enceladus. At 10~Myr we reverted Enceladus to an equilibrium tidal evolution rate and current tidal parameters, evolving the system until 20~Myr or until Mimas-Tethys resonance reached its present state. We also assumed equilibrium tides on Rhea before 10 Myr and accelerated evolution after than (Section \ref{ssec:rhea}), and accelerated Titan throughout, but this had little importance as nether Rhea not Titan encountered any resonances.  

\begin{deluxetable*}{lcccc}
\tablenum{1}
\tablecaption{Initial conditions and tidal parameters for the simulations shown in Fig. \ref{bern}. The slashes separate the values relevant for the two simulations (left-hand side is listed first). For parameters that were changed during the simulation, vertical lines separate values used during 0-3~Myr, 3-10~Myr and after 10~Myr, respectively. Relative tidal acceleration refers to a factor by which the Saturn's response of $k_2/Q=(4000)^{-1}$ was enhanced at that moon's frequency. Since $Q=100$ was typically assumed, the last column is usually equal to the tidal love number $k_2$, except for Enceladus during 3-10~Myr period, when $k_2=1$ and $Q < 100$.\label{initial}}
\tablewidth{0pt}
\tablehead{
\colhead{Moon} & \colhead{Eccentricity} & \colhead{Inclination} & \colhead{Relative Tidal} & \colhead{Tidal $k_2/Q$}\\
\colhead{Name} &  \colhead{$e$} & \colhead{ $i$ [$^{\circ}$]} & \colhead{Acceleration} & \colhead{$\times 10^2$}}
\startdata 
Mimas & 0.022 & 0.1 & 1 & 0.001\\
Enceladus & 0.0035 & 0.003 & $10 | 10 | 1$ & $0.01 | 2.7 / 3.3 | 1$\\
Tethys & 0.002/0.003 & 0.98 & 0.33 & 0.02 \\
Dione & 0.005 & 0.03 & 1 & 0.02\\
Rhea & 0.002 & 0.33 & $1 | 1 | 5$ & 0.02 \\
Titan & 0.0305 &  0.33 & 10 & 0.3\\
\enddata
\end{deluxetable*}

\begin{figure*}
\plotone{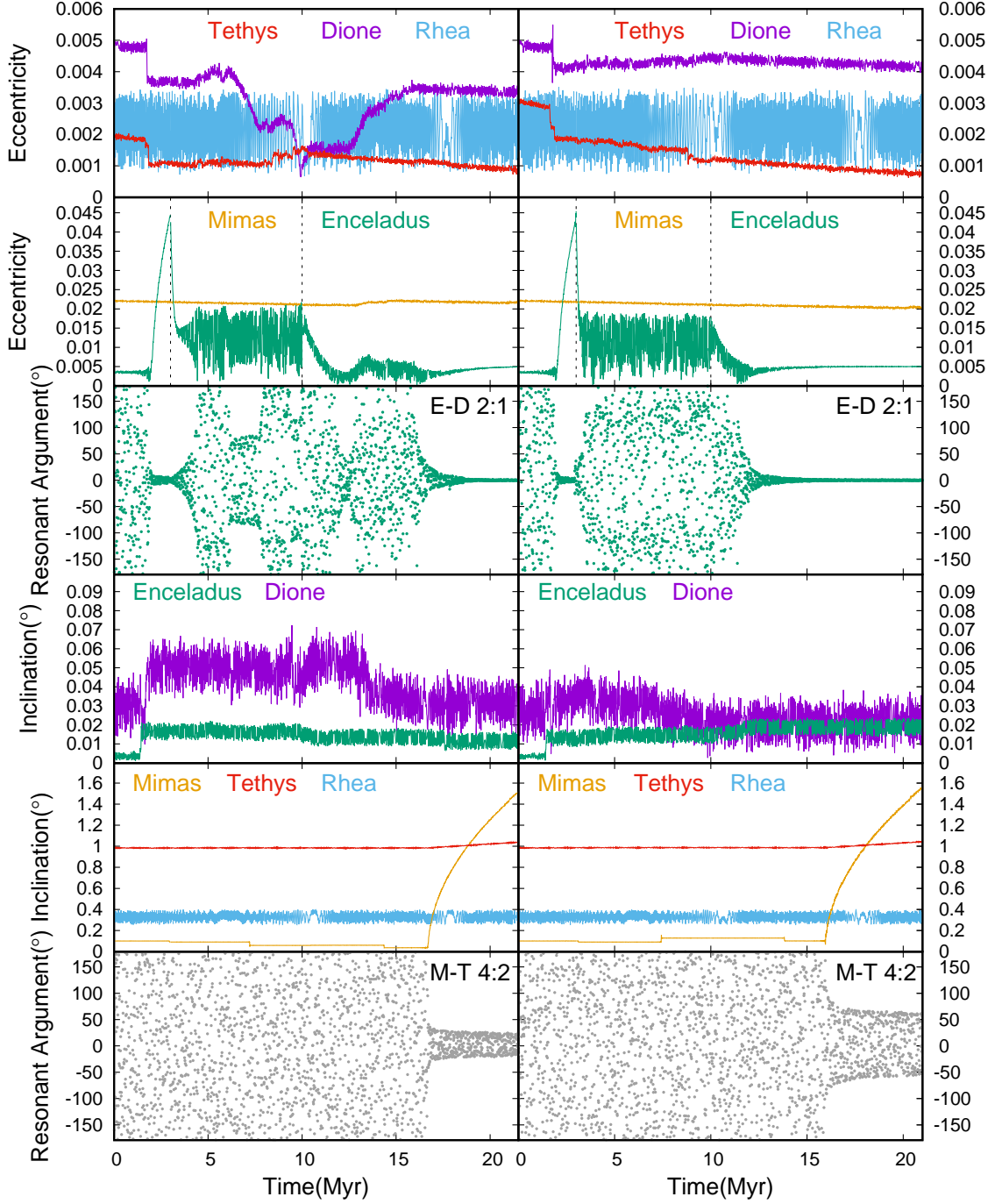}
\caption{Two different simulations (plotted on the left and righ-hand side, respectively) of recent evolution of he inner moons assuming 10x stronger tides acting on Enceladus in the first 10 Myr of the simulation. Most moons evolved using a frequency-independent equilibrium tides with $Q/k_2=4000$ for Saturn. Initial conditions are set at about 20~Myr ago, with the goal of reproducing the current system (Table \ref{initial}). The moons' eccentricities and inclinations were divided into two panels each to clearly show both high and low values. Vertical dashed lines show times (3 and 10~Myr) at which the tidal response and/or acceleration of Enceladus were changed (see Table \ref{initial}). The panels labelled ``E-D 2:1'' plot Enceladus-Dione resonant argument $2\lambda_D - \lambda_E - \varpi_E$, and panels labelled ``M-T 4:2'' plot Mimas-Tethys argument $4\lambda_{\Theta}-2\lambda_M-\Omega_M-\Omega_{\Theta}$.\label{bern}}
\end{figure*}

Figure \ref{bern} shows outcomes of two successful simulations (one from each set), which come close to reproducing the current system. Between two sets of first-stage simulations (20 total), ten were ``successful'', meaning that Enceladus and Dione were in their 2:1 $e_E$ MMR at 3~Myr. The other ten simulations were caught in one of the other sub-resonances of the Enceladus-Dione 2:1 MMR (Fig. \ref{etv_ecc}), typically exciting the eccentricities of Tethys or Dione well above the observed values, or ``fell out'' of the resonance due apparent interaction with secondary resonances. We chose two of the successful simulations for cloning, which was done by sharply changing the tidal properties of Enceladus. The tidal Love number was changed from $k_{2E}=0.01$ to $k_{2E}=1$, approximating melting. Different clone simulations had tidal $Q$ of Enceladus in the 30-39 range. This range was chosen through trial and error, as we found that simulations with $A=(Q_S k_{2E})/(Q_E k_{2S}) < 10$ lead to breaking of the resonance, while those with $A > 10$ settled into bound cycle within the 2:1 $e_E$ sub-resonance (subscript $S$ designates Saturn's properties at Enceladus's frequency). This critical change in the dynamics of resonance with the change in Enceladus's tidal response was first discovered by \citet{mey08}, and our value for critical $A$ agrees with theirs\footnote{As \citet{mey08} explored a more standard case in which the orbital evolution of Dione is not negligible compared to that of Enceladus, $A$ in their paper must be multiplied by $(1-(\dot{a}_D a_E)/(\dot{a}_E a_D))^{-1}$.}. This phase of the evolution of Enceladus, despite seemingly chaotic oscillations, is not stochastic but determined by tidal properties, so all simulations in which Enceladus is dissipative enough (i.e. $A > 10$) will stay at the threshold of the Enceladus-Dione 2:1 $e_E$ resonance indefinitely. 

We assume that Enceladus eventually gets out of the resonant mode and migrates due to ``background'' tidal response of Saturn that is not dependent on frequency. In reality this transition would be gradual as resonant modes are expected produce a smooth (if narrow) profile of tidal response over frequency \citep{ful16}. We could not simulate this without extensively modifying the integrator and, more importantly, greatly increasing the complexity of our model. At 10~Myr all of our clone simulations switch to Enceladus experiencing Saturn's tides with $Q/k_2=4000$ and having its own $k_2/Q=0.01$, approximately what is expected if the current heating is equal to observed and is in equilibrium. Out of twenty cloned simulations only three do not complete the transition to a narrow libration within the resonance consistent with observations (in two simulations Enceladus leaves the resonance, in one it is permanently trapped in a large libration state due to secondary resonance). In some simulations the transition is smooth like in the right-hand side of Fig. \ref{bern}, while in the others secondary resonances are encountered during the evolution, leading to temporary increases in the libration amplitude (as in the left-hand side of Fig. \ref{bern}).

The migration of Mimas, Tethys and Rhea is largely unaffected by the Enceladus-Dione resonance in our simulations. There is some variation of the eccentricity of Mimas when Enceladus encounters secondary resonances within the 2:1 $e_E$ MMR with Dione, but the effect on Mimas is usually minor. The inclination of Mimas is unaffected by any dynamics discussed so far, and passes through a series of kicks due to sub-resonances of the Mimas-Tethys 2:1 commensurability. The subresonances with arguments $2\lambda_{\Theta}-\lambda_M+\varpi_M-2\Omega_M$, $2\lambda_{\Theta}-\lambda_M+\varpi_M-\Omega_M-\Omega_{\Theta}$ (cf. Fig. \ref{3incres}), and $4\lambda_{\Theta}-2\lambda_M-2\Omega_M$ (the pure $i^2_M$ term, Fig. \ref{mimteth}) are encountered at about 3~Myr, 7~Myr and 14~Myr, respectively. Since the latter two kicks (which are larger) always happen after the simulations were cloned at 3~Myr, we can consider clones of the original two simulations to be practically independent when it comes to the inclination of Mimas. Out of eighteen clones which stayed in the Enceladus-Dione 2:1 MMR, nine were captured into the current Mimas-Tethys 4:2 $i_M i_{\Theta}$ sub-resonance (bottom panels in Fig. \ref{bern}). The remaining clones either experienced capture into the $i_M^2$ harmonic or passed the whole forest of sub-resonances without capture. Therefore we can say that both observed resonances (Mimas-Tethys 4:2 $i_Mi_{\Theta}$ and Enceladus-Dione 2:1 $e_E$) have high probability (about 50\%) when using these initial conditions and assumptions about Enceladus's tidal evolution and internal dissipation. While we can quantify the probabilities of different stochastic outcomes of resonant dynamics, we are currently not able to assess the {\it a priori} probability of our initial conditions and assumptions about Saturn's tidal response. Note that our preferred timeline requires Enceladus to settle into the present quiescent state before Mimas acquires high inclination through its resonance with Dione. The current Mimas-Tethys resonance is very fragile and we often see it broken if it coincides with Enceladus having very high and/or chaotic eccentricity. 

\begin{figure*}
\epsscale{.8}
\plotone{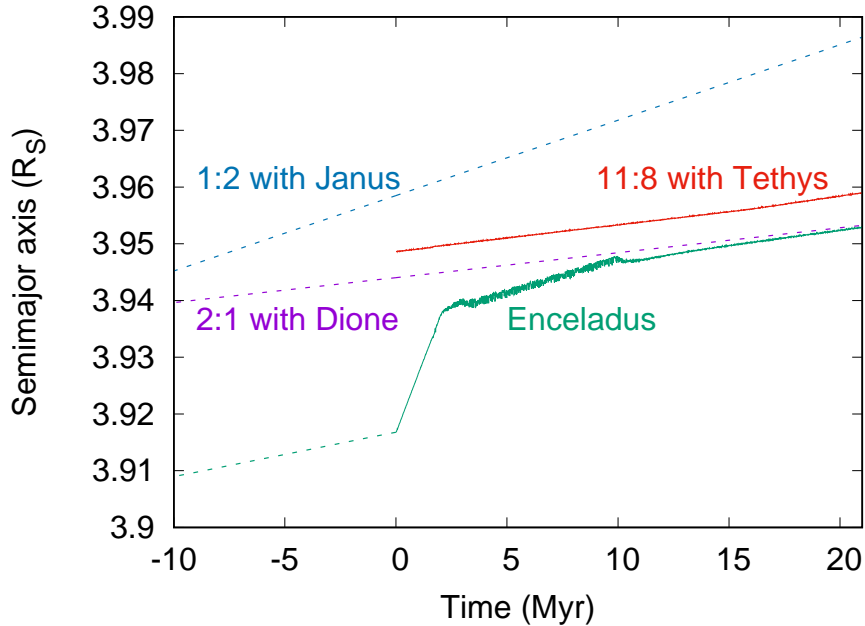}
\caption{Semimajor axis of Enceladus (solid green line) in the right-hand simulation in Fig. \ref{bern}, compared to locations of mean-motion resonances with other moons. Solid lines are from this numerical simulation, while dashed lines are analytical estimates. The blue dashed line shows the location with the 1:2 resonance with Janus or its progenitor, assuming orbital evolution due ring torques \citep{taj17}. The solid red line is the location of the 11:8 Enceladus-Tethys resonance in our simulation. The purple dashed line extrapolates the location of 2:1 resonance with Dione into the past assuming equilibrium tides only. The green dashed line extrapolates the semimajor axis of Enceladus before the simulation starts assuming equilibrium tides. \label{semimajor}}
\end{figure*}

Other parameters of the system at the end of simulations in Fig. \ref{bern} are mostly consistent with the present state. Mimas-Tethys resonance in every case has a large libration amplitude that is still well short of the observed one ($93^{\circ}$), but we expect even very minor subsequent perturbations to modify this quantity (we ignored all moons smaller than Mimas in these runs). Inclinations of Enceladus and Dione are somewhat stochastic but in the range that includes the observed values ($i_E=0.008^{\circ}$, $i_D=0.02^{\circ}$), as is the eccentricity of Dione (probably the most variable quantity in our simulations; currently $e_D=0.002$). For both Mimas and Tethys inclinations are determined by the initial conditions and their mutual resonance, and the eccentricities depend primarily on the initial values and tidal dissipation, without much stochasticity. However, the eccentricity of Rhea is one quantity that our simulation cannot explain, as we have set it much higher than observed in order to avoid three-body argument of the Enceladus-Dione MMR (Fig. \ref{etv_ecc}), but the tidal dissipation we assume for Rhea cannot subsequently modify that eccentricity. This discrepancy may tell us something about the recent dynamics of Rhea, as we discuss in the next Section.

\section{Discussion and Conclusions}\label{discon}

\subsection{Damping of Inclination in Enceladus?}\label{damp}

Much of our reasoning that resulted in the model of recent evolution presented in Fig. \ref{bern} is driven by the survival of the low inclination of Enceladus. In order for Enceladus's inclination to stay so low, both the capture into the 4:2 Enceladus-Dione MMR (Sections \ref{ssec:ed42} and \ref{fast_ed42}) and crossing of the 11:8 Enceladus-Tethys MMR (Section \ref{ssec:et118}) must be avoided. Can we be confident that the inclination of Enceladus was not actually higher in the past and then damped by tides? 

The simplest way to estimate tidal damping of inclination within Enceladus is assuming a homogeneous Enceladus that can be described by a single tidal love number $k_2$ and a tidal quality factor $Q$. If we assume that the eccentricity of Enceladus is currently in equilibrium between excitation by resonance with Dione and damping by tides which produces the observed heating \citep{how11}, we get $k_2/Q \approx 0.01$ \citep{mey07, lai12}. This corresponds to an eccentricity-damping timescale of about 0.5~Myr, making Enceladus exceptionally dissipative. The timescale for damping of inclination is longer by the factor $7 (\sin{i}/\sin{\theta})^2$, where $\theta$ is the forced obliquity of the moon. The forced obliquity of Enceladus has been modeled by \citet{bal16} and they found that $\theta \le 4\times 10^{-4 \circ}$. This would make the $(\sin{i}/\sin{\theta}) \ge 20$, making the timescale for damping of Enceladus's inclination over 1~Gyr, clearly too slow to affect dynamics in the timeframe we consider.

Another mechanism for inclination damping would be resonant response of a global ocean \citep{tyl08, tyl11}. Given that Enceladus likely possesses a sub-surface ocean, this possibility must be addressed. \citet{che14} analysed in-depth different mechanisms of tidal dissipation, and found that at Enceladus's present obliquity the heating due to obliquity tides in the ocean is more than three orders of magnitude lower than non-resonant obliquity tides due to whole-body response by Enceladus. This also implies that the inclination damping timescale due to resonant ocean tides is longer than $10^3$~Gyr. One important finding by \citet{che14} about resonant tides is that their power depends as a cube of obliquity (as opposed to the square of obliquity in non-resonant tides), meaning that the timescale for inclination damping is inversely proportional to obliquity. However, even if the inclination of Enceladus were about a degree (cf. Fig. \ref{etv_inc}) and forced obliquity were $\theta \approx 0.05^{^{\circ}}$, resonant inclination damping would still be an order of magnitude slower than non-resonant obliquity tides. Finally, even if the parameters used by \citet{che14} were not to apply (for some unforeseen reason) to Enceladus on a high inclination orbit, their calculations clearly demonstrate that the inclination of Enceladus cannot be brought to its present low value by resonant ocean tides.

The above discussion assumed that Enceladus is in Cassini state 1, in which the forced obliquity due to inclinations is low, and that dissipation within other moons does not affect the evolution of Enceladus's inclination. However, it is in principle possible that the obliquity of a moon could be excited by a spin-orbit resonance. This would be equivalent to the excitation of obliquity of Saturn itself by a secular resonance between the axial precession of Saturn and nodal precession of Neptune's orbit \citep{war04, ham04}. Recently \citet{cuk20} suggested that Uranian moon Oberon may have in the past been caught in a similar resonance with the orbit of Umbriel, and that tidal dissipation within Oberon may have damped the inclination of Umbriel. Using the results of \citet{che14}, we surveyed the precession frequencies in the Saturnian system and did not find any candidates for such a resonance, although we cannot exclude this possibility due to uncertainties in the moons' shapes and gravity fields. There are however additional arguments against the relevance of such resonances to Enceladus. Most importantly, dynamics of the Saturnian satellite system is dominated by Saturn's oblateness, and the mutual perturbations by the moons are less important than in the Uranian system. Therefore there is little or no coupling between the precession frequencies of different moons, and the relatively low mass Enceladus is particularly unlikely to have a noticeable effect on the rotational dynamics of the larger moons. Furthermore, \citet{cuk20} found that damping of inclination through a spin-orbit resonance can operate only until the relevant orbital inclination damps below the level at which resonance can be maintained, and this level is likely to be higher than the very low current $i_E=0.008^{\circ}$. Therefore we conclude that spin-orbit resonances are vey unlikely to have damped the inclination of Enceladus. 

\subsection{Implications of the Moons' Initial Conditions}

{\bf Mimas} Our model requires that the eccentricity of Mimas predates the establishment of the current Mimas-Tethys and Enceladus-Dione resonances. The origin of this eccentricity may lie in the past 3:1 Mimas-Dione MMR \citep[cf. ][]{mey08, cuk22}, or some other still unidentified, possibly 3-body, resonance. The implication is that the damping of Mimas's inclination was limited and therefore that Mimas is unlikely to posses an internal ocean \citep{rho22}. A new finding in our paper is that Mimas must have had a $\approx 0.1^{\circ}$ inclination before encountering 2:1 MMR with Tethys. The origin of this inclination is harder to explain, as most three-body resonances do not affect inclination, and 3:1 MMR crossing with Dione produces a smaller inclination ``kick'' \citep{cuk22}. One possibility that needs investigation is whether Dione was more inclined at the time of Mimas-Dione 3:1 MMR crossing, and that would require producing a self-consistent model of the system's evolution further than 20~Myr into the past.

{\bf Enceladus} We find that the eccentricity of Enceladus was already excited ($e \le 0.005$) before encountering the 2:1 MMR with Dione, but its inclination was not. The most likely source of this eccentricity are three-body resonances, either isolated or as part of the 5:3 Dione-Rhea MMR crossing. An apparently pristine inclination of Enceladus implies that it did not encounter any major (1st or 2nd order) two-body resonances with mid-sized moons prior to the 2:1 MMR with Dione.

{\bf Tethys} Our model requires Tethys to have a somewhat higher eccentricity ($e_{\Theta}=0.002-0.003$) 20 Myr ago, and almost its present large inclination. The eccentricity of Tethys poses no challenges, as a past higher eccentricity was suggested on both geophysical \citep{che08} and dynamical \citep{cuk16} grounds, and tidal dissipation is likely to subsequently produce Tethys's present low eccentricity. The one-degree inclination of Tethys was always known to have to predate the Mimas-Tethys 4:2 resonance \citep{all69, sin72}, but its origin was never explained until \citet{cuk16} proposed a Tethys-Dione secular resonance closely following (and dynamically related to) the Dione-Rhea 5:3 MMR crossing. Unless another mechanism of exciting Tethys's inclination is found, our initial conditions effectively require that a passage through the Dione-Rhea 5:3 had happened in the past.  

{\bf Dione} We start our simulations with Dione that has excited eccentricity ($e_D \approx 0.005$) but very low inclination. This is generally consistent with a past secular resonance with Tethys \citep{cuk16}, in which Dione ``passed'' its inclination and part of its eccentricity (acquired in a resonance with Rhea) to Tethys. The secular resonance is broken when either the eccentricity or inclination of Dione reach very low values (e.g. $i_D<0.1^{\circ}$ for inclination). Therefore, a substantial eccentricity is to be expected to survive if the inclination is very low, implying that the resonance was broken by depletion of Dione's inclination.

{\bf Rhea} At the start of our simulation $e_R=0.002$, ten times in excess of Rhea's current free eccentricity $e_R=2 \times 10^{-4}$.\footnote{The total eccentricity of Rhea is larger due to a dominant term forced by Titan which is not relevant for dynamics discussed here.} Rhea's inclination at the start of our simulation is the same as now ($i_R=0.33^{\circ}$) and does not change in the course of it. This substantial inclination is exactly what is expected from a past crossing of the Dione-Rhea 5:3 MMR. This dynamical mechanism is also expected to produce a comparable eccentricity of Rhea, broadly consistent with our initial conditions, but not the observed values. The discrepancy between theoretical expectations and the actual value here is significant, and requires so-far unknown dynamical mechanism of lowering Rhea's eccentricity. We plan to address this issue in the near future using an integrator that resolves the rotational dynamics of Rhea and search for any additional dynamical features. 

\subsection{Width and Distribution of Peaks in Tidal Dissipation}

Our work is decidedly semi-empirical in design, as we acknowledge the importance of highly variable response of Saturn to tidal forces of different moons. As we have shown in this paper, both equilibrium tides and resonance-lock-only models that were used so far fail to fully explain the system's dynamics. Apart from the general lack of information on the source of Saturn's dissipation, we tried to keep the number of free parameters to a minimum, leading to our decision to change the evolution rates abruptly. Our decision not to model the interior evolution of Enceladus but to periodically adjust its tidal parameters ``by hand'' was dictated by our own technical limitations and and we hope that in the future there will be integrated models that fully model both orbital dynamics and interior evolution of the moons.

Regardless of the nature and evolution of resonant modes, it is clear that these are resonant phenomena, and therefore must exhibit a spike-like profile against frequency (possibly Lorentzian or similar). As apparent from Fig. \ref{semimajor}, we assumed that the width of the normal mode affecting Enceladus to be about 1\% of its semimajor axis. We could in principle restrict the very fast evolution to a narrower interval that includes the initial encounter with the Enceladus-Dione 2:1 MMR, but that would require that the resonant mode and the resonance with Dione were encountered at the exact same time due to an unlikely coincidence.  Furthermore, if Rhea is only passing though a resonant mode, this mode cannot be too narrow if we were to observe this transient phenomenon. These features in frequency space appear much wider than the resonant modes proposed by \citet{ful16} based on the theory of tidal response in stars and giant planets.  We also find that the tidal response at Tethys's frequency must be lower than what appears to be ``background'' rate, so this frequency dependence is not restricted to peaks in dissipation, but also has local minima (``troughs''). Apart from being relatively wide, resonant modes cannot be too few and far in between if they were recently encountered by both Enceladus and Rhea as we propose. 

The determination of a consensus result for the current rate of tidal evolution of Titan is of great importance, as it will give us a major indication if resonance locking is present or not. Of all the moons of Saturn that raise significant tides (Mimas-Titan), Titan is most likely to be in resonant lock, as it has the slowest (equilibrium) tidal evolution and is the least likely to have been recently re-accreted in some kind of late cataclysm. If Titan is not in resonance lock, it is possible that the resonant modes are moving inward (in terms of semimajor axis at which they are encountered), making is somewhat less surprising that encounters between moons and modes appears to common. Still, two moons encountering resonant modes within the last few tens of Myr appears unlikely unless modes are very numerous, or their distribution is correlated with that of moons. The last possibility may indicate that the moons may have re-accreted close to the modes, either due to where the last generation of satellites was before the assumed instability, or for some other reason. The only certainty is that the tidal evolution of the Saturnian system holds yet more surprises for us.

\subsection{Summary}

In this work we tried to consider all of the available constraints on the recent (last few tens of Myr) and current orbital evolution of major Saturnian satellites. We find that no single mechanism of orbital evolution proposed so far (including frequency-independent equilibrium tides and the evolution through resonance-locking) can explain the orbits of these moons. Strong equilibrium tides can explain the existence of the observed resonances (Mimas-Tethys and Enceladus-Dione) and the current heating rate of Enceladus, while resonant modes are necessary to explain the current dynamics of Rhea and the original capture of Enceladus into the resonance with Dione. Additionally, the evolution of Tethys needs to be slower than that of other moons, implying ``troughs'' as well as ``peaks'' in response as a function of frequency. 

In order to successfully reproduce the encounter of Enceladus with the 2:1 resonance with Dione we require a {\it passage through} a resonant mode, rather than {\it locking to} a resonant mode. This is reasonable if the resonant modes in the inner system are evolving more slowly than those at larger distances, as originally predicted \citet{ful16}, but would not work in the context of inertial waves \citep{lai20}. Alternatively, if Titan is currently not locked to a resonant mode as the results of \citet{jac22} suggest, it is also possible that resonant modes move inward, and in that case the moons could only temporarily cross the modes, rather than become locked to them.     

Given the complexity and uncertainties of the tidal evolution rates that not only vary from moon to moon but also over time, it is difficult to reach firm conclusions about the age of the system. However, given the amount of dynamical excitation that the inner moons (especially Mimas and Enceladus) may have experienced in the last 20 Myr, it is difficult to envision this system of relatively ``dynamically cold'' satellites evolve this way for hundreds of Myrs, let alone multiple Gyrs. We hope that more precise future determinations of the current orbital evolution rates of the Saturnian moons (based on astrometry or spacecraft data) will be able to confirm or falsify our model of their recent evolution.  

\begin{acknowledgments}
This work was supported by NASA Solar System Workings Program awards 80NSSC19K0544 (to M\'C and MEM) and 80NSSC22K0979 (to M\'C). We would like to thank Jim Fuller, Valery Lainey, Bob Jacobson, and Francis Nimmo for very insightful discussions. We also thank the International Space Science Institute in Bern for organizing an extremely useful workshop in the evolution of the Saturnian system (May 2022). We wish to thank two anonymous reviewers whose comments greatly improved the paper.  
\end{acknowledgments}

\bibliography{refs_saturn}{}
\bibliographystyle{aasjournal}

\end{document}